\documentstyle[aps,prl]{revtex}
       \def\Journal#1#2#3#4{{#1} {\bf #2}, #3 (#4)}
       \def\PLB{{Phys. Lett.}  B}
       \def\PRL{{Phys. Rev. Lett.}}
       \def\PRB{{Phys. Rev.} B}
       \def\ZPB{{Z. Phys.} B}
       \def\et{{\it et al.}}
       \def\etws{{\it et al. }}
\begin{document}
\input epsf.sty
\twocolumn[\hsize\textwidth\columnwidth\hsize\csname %
@twocolumnfalse\endcsname
\draft
\widetext

\title{Spin Correlations in an Isotropic Spin-5/2 Two-Dimensional 
Antiferromagnet}

\author{R. L. Leheny, R. J. Christianson, and R. J. Birgeneau}
\address{Department of Physics, Massachusetts Institute of Technology,
Cambridge, MA 02139, USA}

\author{R. W. Erwin}
\address{Reactor Radiation Division, NIST, Gaithersburg, MD 20899, USA}

\date{\today}
\maketitle

\begin{abstract}
We report a neutron scattering study of the spin correlations for the spin 5/2,
two-dimensional antiferromagnet Rb$_2$MnF$_4$ in an external magnetic field.  
Choosing fields near the system's bicritical point, we tune the effective 
anisotropy in the spin interaction to zero, constructing an ideal $S=5/2$ 
Heisenberg system.  The correlation length and structure factor amplitude are 
closely described by the semiclassical theory of Cuccoli {\it et al.} over a 
broad temperature range but show no indication of approaching the low-temperature
renormalized classical regime of the quantum non-linear sigma model.

\end{abstract}
\


\phantom{.}
]
\narrowtext

Magnetic systems with reduced dimensionality have provided a basis for numerous
insights into the varying roles quantum and thermal fluctuations play in driving
phase transitions.  The two dimensional quantum 
Heisenberg antiferromagnet (2DQHA) represents
a particularly important example of such a system due to the possible connections
between it and the origins of superconductivity in the layered 
copper oxides~\cite{highTc}.  
Specifically, the parent compounds to the 
high T$_c$ superconductors, such as La$_2$CuO$_4$, 
are good examples of 2DQHA's on a square lattice with spin, $S=1/2$, and the
magnetism persists into the superconducting state~\cite{highTc}.
As a consequence of this connection, focus on the 2DQHA has intensified in recent
years, and the combined efforts of experiment, theory, and simulation have
yielded an increasingly cohesive picture of its behavior.

In mapping the 2DQHA onto the quantum nonlinear sigma model (QNL$\sigma$M), 
Chakravarty, Halperin, and Nelson
have produced an effective field theory for the system which
predicts renormalized classical behavior
at low temperatures.
Calculations from this theory (labelled 
CHN-HN~\cite{chn}), describe a quasiexponential 
growth in the correlation length, $\xi$, 
with inverse temperature that matches  
measurements on $S=1/2$ systems~\cite{greven} 
with good absolute agreement over a remarkably large
temperature range.  However, high temperature results from 
systems with $S>1/2$ vary markedly from the CHN-HN 
predictions~\cite{greven,nakajima,lee,fulton}. 
These discrepancies have led to questions  
regarding the applicability of the effective field theory
in the temperature range probed in these 
experiments
and have motivated new approaches to describing 
the 2DQHA at high temperatures.  For example,
Elstner \et~\cite{elstner} have applied series expansion techniques 
to characterize the onset of correlations with good accuracy.
In addition, Cuccoli \et~\cite{cuccoli} have recently proposed a 
semiclassical theory, the pure quantum self consistent harmonic
approximation (PQSCHA).  This theory calculates
thermodynamic quantities, such as  $\xi$, with respect to their
classical values, with the effects of quantum fluctuations 
entering as a renormalization of temperature.
While this semiclassical picture should be valid 
only at high temperature for $S=1/2$, the authors predict that the 
theory should work over a broad range of temperatures
for $S>1/2$~\cite{cuccoli}.

A major obstacle to forming a unified picture of the 2DQHA,
in which the relative merits of these theoretical approaches
can be distinguished, 
has been the absence of experimental 
systems with $S>1/2$ that model the 2DQHA 
over a broad range of length scales and, concomitantly, temperatures.
In particular, the materials that have been studied, which include
K$_2$NiF$_4$~\cite{greven}  and La$_2$NiO$_4$~\cite{nakajima}
with $S=1$ and Rb$_2$MnF$_4$~\cite{lee} and KFeF$_4$~\cite{fulton} with $S=5/2$, 
possess appreciable Ising anisotropies that generate crossovers from 2D
Heisenberg to 2D Ising behavior at quite short length scales.
For example, in both Rb$_2$MnF$_4$ and KFeF$_4$ the crossover to Ising behavior 
is first apparent at $\xi/a \simeq 6$, where $a$ is the lattice spacing.    
In contrast,  the 2D $S=1/2$ material Sr$_2$CuO$_2$Cl$_2$ 
exhibits model 2DQHA behavior in its spin correlations 
for length scales up to $\xi/a \simeq 200$~\cite{greven}.
As spin number is the central parameter controlling the strength of quantum effects, 
measurements over a broad range of $\xi$ away 
from the extreme quantum limit, $S=1/2$, 
are crucial to enhance our understanding of the 2DQHA.  

In this paper we present a study of Rb$_2$MnF$_4$ under conditions
in which its spin space anisotropy is effectively reduced to zero, thus allowing us 
to characterize the Heisenberg behavior of this $S=5/2$ system over a 
range of length scales which rivals those probed in
the experiments on the $S=1/2$ systems. 
Rb$_2$MnF$_4$ has the 
K$_2$NiF$_4$ crystal structure, in which magnetic
planes of MnF$_2$ form a square lattice with nearest neighbor 
spacing, $a=4.215 \AA$ at 5 K. 
The Mn$^{2+}$ ions 
($S=5/2$) interact antiferromagnetically
with nearest neighbors on the lattice with an isotropic 
exchange, $J=7.36$ K, plus 
a uniaxial, dipolar field along the \^{c} axis (perpendicular to the planes) 
of strength, $g\mu_BH_A \approx 0.005J$~\cite{dewijn}.  

To reduce the effective anisotropy we place Rb$_2$MnF$_4$
in an external magnetic field, H, parallel to  \^{c}.
As studied in detail by Cowley \et~\cite{cowley}, 
the phase diagram for Rb$_2$MnF$_4$ 
in a uniform magnetic field contains a 
spin-flop transition between the low-field Ising phase
and a high-field XY phase.  Results from Ref.~\cite{cowley}
along with new 
measurements, shown in Figure 1, reveal the  
$T=0$, 2D bicritical point, where the boundaries between these
phases and the paramagnetic phase meet, 
at $H_B = 5.30\pm0.12$ T.
The effective anisotropy, $g$, in the
spin interaction varies with field  
as $g \sim H^{2}-H_{B}^2 -CT$~\cite{cowley}, with $g<0$ denoting
easy axis anisotropy and $g>0$ denoting easy plane.
Thus, tuning the field at a given temperature provides
a powerful mechanism for controlling the symmetry
and magnitude of the effective anisotropy.
In particular, 
for the fields at which $g=0$ 
(the dashed line in Fig.\ 1), 
the effective anisotropy 
vanishes, and all spin components are
critical.  In our experiment we have carefully
chosen fields to match this condition
as closely as possible, thus avoiding 
a crossover to a lower symmetry transition.  
Specifically, we obtain $H_{B}^2$ and $C$
from the spin-flop line at low temperature ($T<25$ K) and
extrapolate this line into the paramagnetic region to 
determine the appropriate fields.  This 
strategy has allowed us to track the Heisenberg behavior in 
Rb$_2$MnF$_4$ to significantly lower temperatures 
and, concomitantly, larger length scales
than otherwise possible.
 
We have characterized the instantaneous magnetic correlations 
of this Heisenberg system by performing 2-axis, 
energy-integrating neutron scattering measurements 
with the BT9 spectrometer at the NIST Center for Neutron Research.  
To accomodate the relative orientations of the spectrometer axes and the external
field, we oriented the crystal such that  \^{c} was 
perpendicular to the scattering plane. 
To guarantee that the quasielastic approximation
was satisfied, we employed a large incident neutron energy, $E_{in}=100$ meV\@. 
We created the 100 meV incident beam using the (0,0,4) reflection from a
pyrolytic graphite monochromator.
The collimator sequence for the spectrometer was 40'-13'-S-10', and
a sapphire filter after the monochromator
removed higher order contamination.
We model the static structure factor, $S({\bf Q_{\rm 2D}})$, 
where $\bf Q_{\rm 2D}$ is the momentum transfer in the magnetic planes,  
with a simple Lorentzian lineshape 
 
\[I({\bf Q_{\rm 2D}}) \propto   S({\bf Q_{\rm 2D}}) = \frac{S_0}
{(1+{\bf q_{\rm 2D}}^2\xi^2)},~~~~~~~~(1) 
\]
where ${\bf q_{\rm 2D}}$ is the wave vector 
measured from the antiferromagnetic zone 
center, ${\bf q_{\rm 2D}} = {\bf Q_{\rm 2D}}-(1/2,1/2,0)$. 
Figure 2 displays scattering profiles
from the 2-axis scans along the direction $(K/2,-K/2,0)$
at three representative temperatures.
We fit the scattering intensity to Eq.\ 1 convolved with the
spectrometer resolution function together with a sloping
background.  The fit results, shown with solid lines in Fig.\ 2,
describe the data very well at all temperatures.
In addition, to test the validity of the 
quasielastic approximation, we have characterized the
shape of the dynamic structure factor, $S({\bf Q_{\rm 2D}},\omega)$,
in energy transfer, $\omega$, by performing
additional 3-axis measurements scanning $\omega$ at ${\bf q_{\rm 2D}}=0$.
A Lorentzian lineshape in $\omega$ fits these 3-axis scans well.
An analysis of the 2-axis scans which accounts explicitly for 
the shape of the dynamic structure factor gives results for $\xi$ and $S_0$ that 
agree with those obtained assuming the 
quasielastic approximation to within 5\% at all temperatures,
thus confirming the validity of the approximation for our 
measurements.

Figure 3a shows $\xi/a$ extracted from 
the fits on a log scale versus $JS(S+1)/T$.  We 
choose $JS(S+1)$ rather than $JS^2$ for the
energy scale following the observation by Elstner \et~\cite{elstner} that $\xi$ 
depends on S primarily through $JS(S+1)$ at high temperature for $S>3/2$.    
Also plotted are the results from Lee \et~\cite{lee}
for Rb$_2$MnF$_4$ in zero magnetic field. 
The good agreement between the two
measurements in the temperature region in which they overlap confirms that 
our method to determine the instantaneous spin correlations is accurate.
In addition, as the plot illustrates, by reducing the effective 
anisotropy to zero we have been able to extend significantly the temperature
range of the Heisenberg behavior.  At our lowest temperature
$\xi/a$ exceeds 100, rivaling the experimental range for the best
$S=1/2$ Heisenberg systems.  

The solid line in Fig.\ 3a is the prediction for $\xi/a$ 
from the PQSCHA~\cite{cuccoli}.  
As the figure demonstrates, the calculated behavior 
from the semiclassical theory 
agrees closely with the measured correlation 
lengths over the full experimental range. 
As mentioned above, the PQSCHA calculates $\xi$ with
respect to its value for the classical Heisenberg model.
For reference, the dashed-dotted line shows the correlation length for 
the classical model, $\xi_{CL}$, as
obtained from simulation~\cite{elstner,kim}.  
At high temperature ($\xi/a<5$),
$\xi$ closely approximates
$\xi_{CL}$ when the energy scale is taken as $JS(S+1)$.  
Sokol \et~\cite{sokol} 
have labelled this temperature region the `classical
scaling regime'.  However,
as the figure demonstrates, with decreasing temperature
quantum effects become increasingly important,
and this simple $S(S+1)$ scaling breaks down.  The
renormalization of temperature that the PQSCHA provides
accounts for the quantum effects nicely. 

Sokol \et~\cite{sokol} have argued that the $S=5/2$ 2DQHA 
should cross over from classical scaling to renormalized classical
behavior near $T \sim \Lambda$, where $\Lambda$ is the upper bound 
of the spin wave energy spectrum (approximately 5.5 meV at 35 K).  
Thus, we estimate this temperature for our system to be near 60 K 
($JS(S+1)/T \approx 1.1$), roughly where the deviations between
$\xi_{CL}$ and the experimental results in Fig.\ 3a become noticeable.  
However, the low-temperature, renormalized classical prediction 
for the QNL$\sigma$M from 
CHN-HN~\cite{chn}, shown with the dashed line in Fig.\ 3a, deviates 
significantly from the measured correlation 
lengths even for $\xi/a$ exceeding 10$^2$. 
This result contrasts with those for the $S=1/2$ systems which show good
agreement with the theory over length scales down to $\xi/a \sim 1$.  
However, recent quantum Monte Carlo simulations combined with finite size 
scaling have indicated that this experimental agreement is
in part accidental due to a cancellation between finite 
lattice effects and higher order corrections 
to the QNL$\sigma$M~\cite{beard}.  In addition, Elstner \et\ 
have argued that the onset of the renormalized
classsical behavior described by the QNL$\sigma$M 
should occur at progressively higher $\xi$ for larger $S$~\cite{elstner}.
Our measurements on this $S=5/2$ 2DQHA, showing deviations
from the projected asymptotic behavior even as the measurements extend well below
the classical scaling regime, support the view of an extended 
crossover region between the classical and renormalized classical scaling regimes.

In Figure  3b we display $S_0$, the amplitude of the static
structure factor at the ordering
wave vector, extracted from the measurements.  The zero field 
results~\cite{lee}, shown with open triangles, have been placed on an absolute scale
by setting the high temperature limit to $S(S+1)/3$.  Our results, placed on the 
same scale, extend the data smoothly to lower temperature. 
The solid line in Fig.\ 3b is the prediction from the PQSCHA
with no adjustable parameters.  As 
with $\xi$, the PQSCHA predicts $S_0$ accurately over a broad
temperature range.

For both the classical~\cite{T2classical} and quantum~\cite{chn} 2DQHA's, 
low-temperature theories predict 
$S_0/\xi^{2} \propto T^2$.  
However, experiments over the range $0.2 < T/2\pi\rho_s < 0.4$ 
for $S=1/2$~\cite{greven} and $S=1$~\cite{nakajima} have found
$S_0/\xi^{2}$ to depend weakly on $T$.   
The inset to Fig.\ 3b shows the results 
for this ratio for Rb$_2$MnF$_4$.   
As the figure illustrates, at high temperature
$S_0/\xi^{2}$ is roughly constant, but at
lower temperature the ratio has a strong    
temperature dependence, qualitatively consistent with
the low-temperature theories.   
The break from temperature independent $S_0/\xi^{2}$ occurs
near $ T/2\pi\rho_s = 0.2$, the lower limit of 
results on other experimental systems.   Thus, 
the measurements to low temperature ($ T/2\pi\rho_s \simeq 0.11$)
in Rb$_2$MnF$_4$ appear to have
captured a crossover from this high temperature regime. 

Finally, we note that the bicritical phase boundaries themselves should 
reflect the 2D correlations.  Specifically, the Ising and XY critical
lines should approach each other tangentially with 
$g \sim T^{-2}\xi^{-2}$~\cite{cowley}.  The solid lines in Fig.\ 1 correspond
to this form with $\xi$ from Fig.\ 3a.  In the fit the $g=0$ line is
fixed by the low temperature spin-flop data, and only the Ising and XY 
amplitudes are adjusted.  Clearly, the agreement is good for temperatures
below $\sim 35 K$.

In conclusion, the presence of a bicritical point in the phase diagram of
Rb$_2$MnF$_4$ has provided a direct
and convenient mechanism for controlling the effective 
anisotropy in the spin interaction of this two-dimensional 
$S=5/2$ antiferromagnet.  
In the present study we have focused on 
conditions in which the effective anisotropy is essentially zero,
allowing us to track the growth of correlations in this
large-spin, two-dimensional
Heisenberg system to correlation lengths exceeding 100 lattice units. 
The success of this approach, the first 
to probe such extended Heisenberg correlations 
in a system with $S>1/2$, suggests that similar studies 
near the bicritical point in tetragonal systems with other spin quantum 
numbers would be fruitful in further extending our understanding of the 2DQHA.
Our measurements for $S=5/2$ demonstrate the success of a semiclassical theory, 
the PQSCHA, over a broad temperature range.  The disagreement between the
measured behavior and the predictions of the effective 
field theory based on the QNL$\sigma$M, even as $\xi$ becomes large,
confirms speculations that the temperature
range applicable to this theory lies well below that accessible in experiment.
However, the growing departure from the 
simple scaling to classical behavior in Fig.\ 3a
along with the marked temperature dependence of $S_0/\xi^{2}$ suggests behavior 
distinct from that seen at high temperature.   
Thus, the experiment reveals an extensive temperature region for the 
crossover between the classical and renormalized classical
regimes.  A quantitative theory for this crossover would be most valuable.
 
We thank R. A. Cowley for his role in conceiving the experiment 
and Y. S. Lee for many helpful discussions.  The research at MIT was supported by 
the NSF under Grant No. DMR97-04532.


\begin{figure}
\centerline{\epsfxsize=3.0in\epsfbox{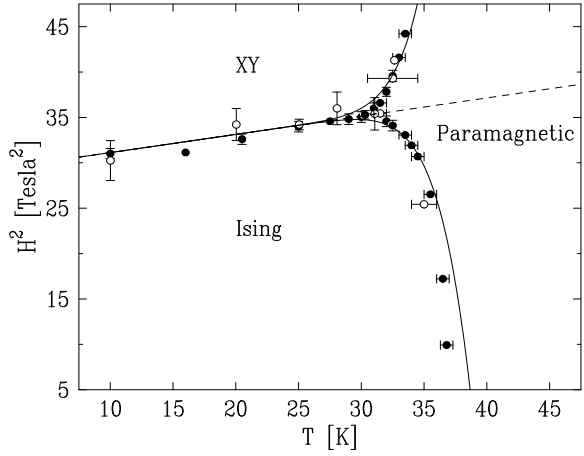}}
\vspace{5 mm}
\caption{Phase diagram for Rb$_2$MnF$_4$ in a magnetic field, H,
parallel to the \^{c} axis.  The solid circles (shifted by 0.15 T) are
from Cowley \etws [10],
and the open circles are from the current experiment.
The solid lines are a fit to the phase boundaries, $g \sim T^{-2}\xi^{-2}$,
using the results for $\xi$ from Fig.\ 3a.
Measurements of the spin correlations at zero effective anisotropy are
taken along the dashed line, which extends from the  
$T=0$, 2D bicritical point
at $H_B =  5.30\pm0.12$ T.}
\label{fig1}
\end{figure}

\begin{figure}
\centerline{\epsfxsize=3.0in\epsfbox{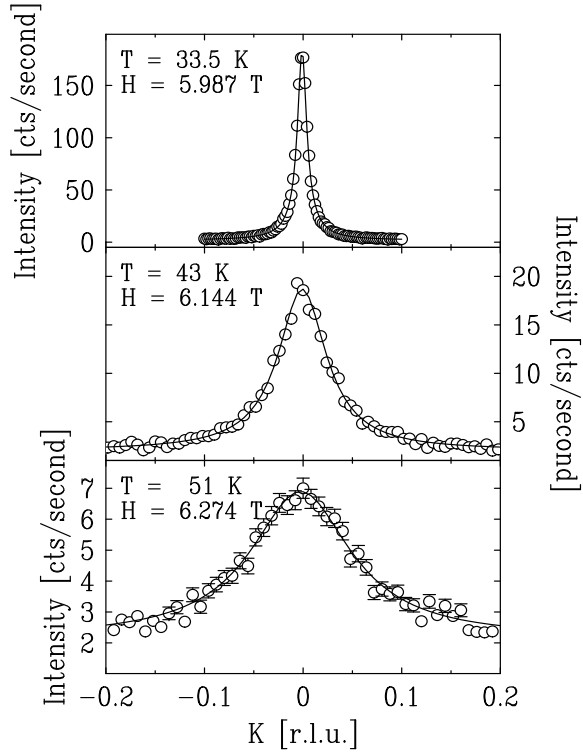}}
\vspace{5 mm}
\caption{2-axis scans along the direction (K/2,-K/2,0)
at field and temperature
values for which the effective anisotropy in the spin interaction
is near zero.  The solid lines are fits to Eq.\ 1 convolved with the
instrumental resolution function.}
\label{fig2}
\end{figure}

\begin{figure}
\centerline{\epsfxsize=3.0in\epsfbox{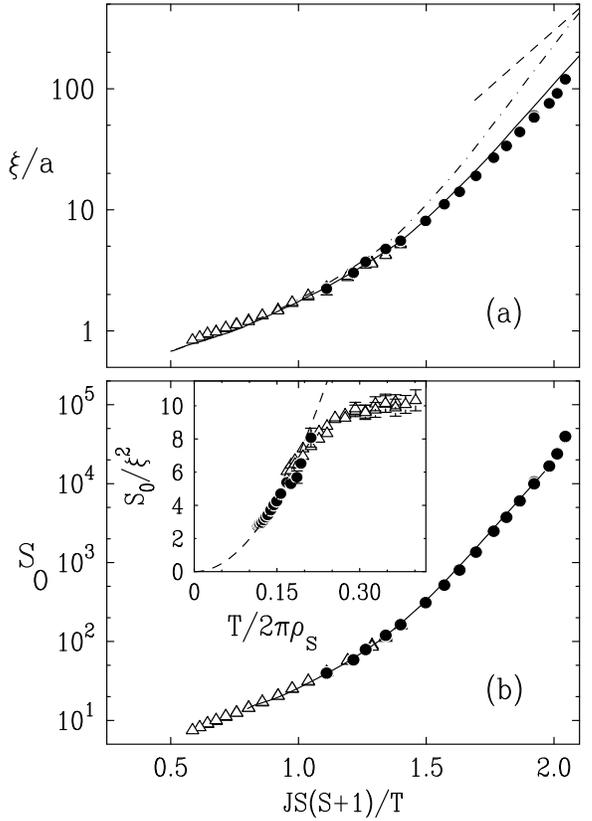}}
\vspace{5 mm}
\caption{(a) The magnetic correlation length, $\xi$, 
in units of the lattice spacing, $a$,
from the present experiment (solid circles) and measured at zero
field in the traditional scattering geometry (open triangles) [5].  
The solid line is the prediction from the semiclassical theory of
Cuccoli \etws [8].  The dashed-dotted line is the correlation
length for the classical Heisenberg model obtained from 
simulation [7,11].
The dashed line is the prediction from low-temperature 
QNL$\sigma$M effective 
field theory of CHN-HN [2].\hspace{0.25in}
(b) $S_0$, the amplitude of the static structure factor at the ordering
wave vector, from the present experiment (solid circles) and  
measured at zero field in the traditional scattering geometry (open 
triangles) [5].  The solid line is
the prediction from the semiclassical theory of Cuccoli \etws [8].  
The inset shows $S_0/\xi^2$
versus temperature.  The dashed line is a fit to the solid circles
assuming $S_0/\xi^2 \sim T^2$.}
\label{fig3}
\end{figure}

\end{document}